\documentclass{ws-p8-50x6-00}

\begin{document}
\def\d0{D\O}
\def\etmis{\mbox{${\hbox{$E$\kern-0.6em\lower-.1ex\hbox{/}}}_T$}}
\def\IEEE{\em IEEE Trans. Nucl. Sci.}

\title{The \d0 Run II Detector and Physics Prospects}

\author{Neeti Parashar}

\address{111 Dana Research Center,Northeastern University, 360 Huntington Avenue, Boston, MA 02115, USA\\
E-mail: neeti@fnal.gov\\
FOR THE \d0 COLLABORATION}


\maketitle

\abstracts{
The \d0 Detector at Fermilab is currently undergoing an extensive upgrade to
participate in the Run II data taking which shall begin on March 1, 2001. The
design of the detector meets the requirements of the high luminosity
environment provided by the accelerator. This paper describes the
upgraded detector subsystems and gives a brief outline of the physics prospects
associated with the upgrade. }

\section{Introduction}
%\subsection{Producing the Hard Copy}\label{subsec:prod}

The \d0 detector
was designed to study
collisions between protons and anti-protons in the Tevatron collider at
Fermilab.
\d0 took its first data run in the period 1992-1996, called Run I.
The successes of Run 1, including the discovery of the top quark, and the
physics potential of high-luminosity running at the Tevatron have additionally motivated the
 present upgrade of the detector. We would now like to pursue a detailed top
quark physics study, search for the Higgs boson and for supersymmetry,
investigate b-physics, and look for physics beyond the Standard Model (SM).

In order to enhance the physics reach of these processes in Run II, the Tevatron also
had to undergo two major upgrades. First, the Tevatron for Run II is
designed to achieve a luminosity of 5x10$^{32}$ cm$^{-2}$s$^{-1}$, which is a factor of
10 more than in Run I. The second upgrade involves a decrease in the bunch
crossing time, which for Run I was 3.5 $\mu s$, while Run II will begin with 396 ns
 and eventually reach 132 ns as the number of bunches is increased.
There is also an increase in the center-of-mass (CM) energy from 1.8 TeV to 2.0
TeV. To take full advantage of the new physics opportunities and to contend
with the high radiation environment and shorter bunch crossing times, an
extensive upgrade of the \d0 detector was undertaken, which is now in its final
stages.

 Figure 1 shows an elevation view of the upgraded detector.
 The upgrade consists of an addition of a solenoid, a tracker
 with silicon and fiber detectors, muon scintillation trigger
 counters, new forward muon system, preshower detectors, readout
 electronics and new trigger and data acquisition systems.

\begin{figure}[p]
%\figurebox{20pc}{15pc}{} % to have a box alone
\epsfxsize=28pc % will enlarge or reduce the postscript figures based on the xsize
%\epsfbox{run2_tracker_annotated2.eps} % postscript image file name
\epsfbox{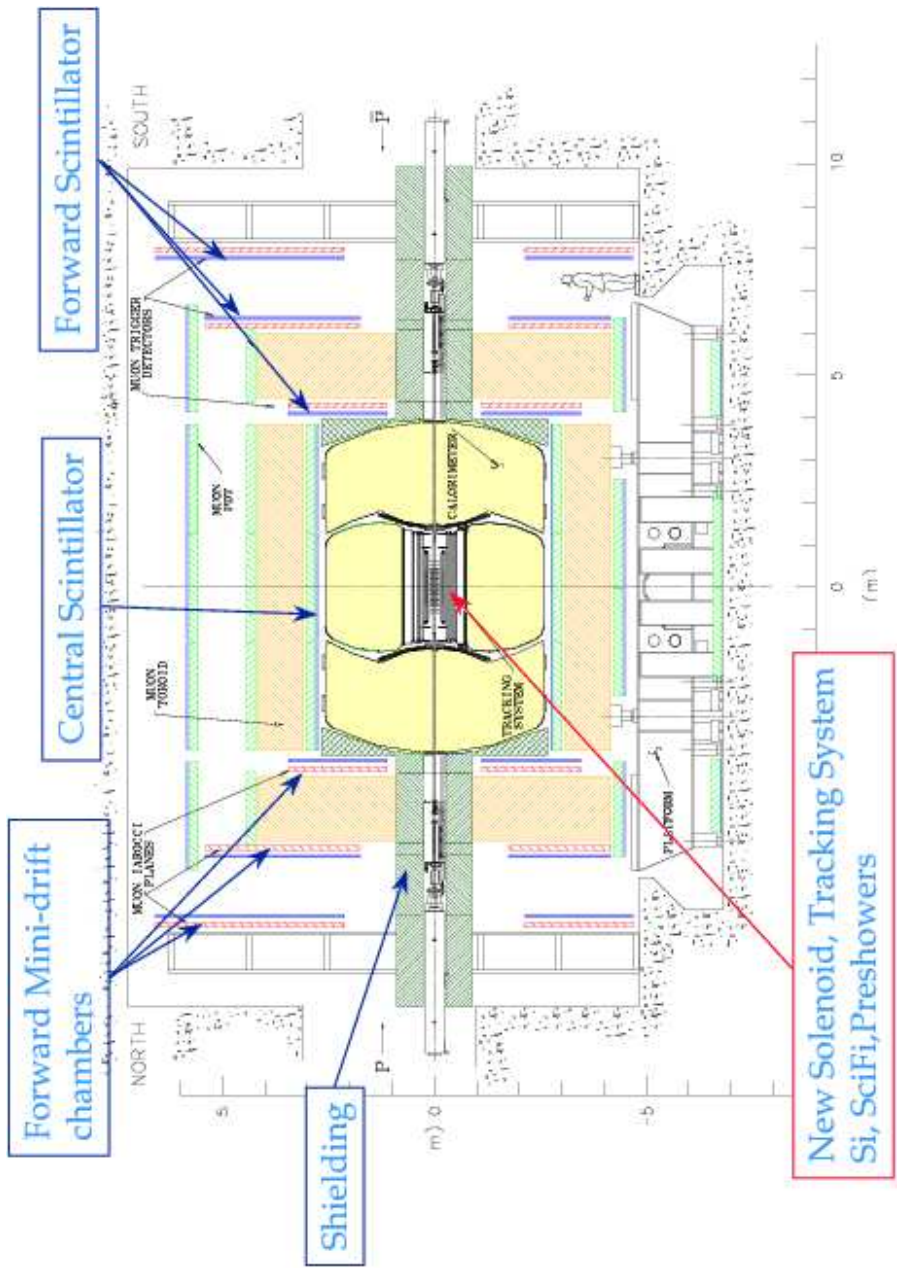} % postscript image file name
\caption{Elevation view of the upgraded \d0 detector for Run II. }
\end{figure}

\section{Tracking}
The upgraded tracking system (Figure 2) consists of an inner silicon vertex
detector,
surrounded by eight superlayers of scintillating fiber tracker. These detectors
 are located inside a 2 Tesla superconducting solenoid, which is surrounded by
 a scintillator based preshower detector. The upgraded tracking system has
been designed to meet several goals:
momentum measurement by the introduction of a solenoidal field; good
electron identification and $e/\pi$ rejection; tracking over a large
range in pseudo-rapidity ($\eta$~$\approx$~$\pm$3); secondary vertex
measurement for identification of $b$-jets from Higgs and top decays
and for $b$-physics; first level tracking trigger; fast detector
response to enable operation with a bunch crossing time of 132~ns; and
radiation hardness.

\subsection{Silicon Microstrip Tracker (SMT)}\label{subsec:wpp}
The silicon tracker is the first set of detectors encountered by particles
emerging from the collision. The detector design consists of interspersed disks
and barrels, based on single and double-sided silicon microstrip detectors,
with a total of 793,000 channels and is radiation hard up to about 2 Mrad.
The SVX IIe chip is
used for readout~\cite{svx2e}. The combination of
small-angle and large-angle stereo provides good pattern recognition
and allows good separation of primary vertices in multiple interaction
events. The expected hit position resolution in $r\phi$ is 10~$\mu m$.

A silicon track trigger preprocessor is being built which will allow the
use of SMT information in the Level 2 trigger.  This will add the capability for
triggering on tracks displaced from the primary vertex, as well as
sharpening the $p_T$ threshold of the Level 2 track trigger and of the
electron and jet triggers at Level 3.

\begin{figure}[t]
%\figurebox{20pc}{15pc}{} % to have a box alone
\epsfxsize=27pc % will enlarge or reduce the postscript figures based on the xsize
\epsfbox{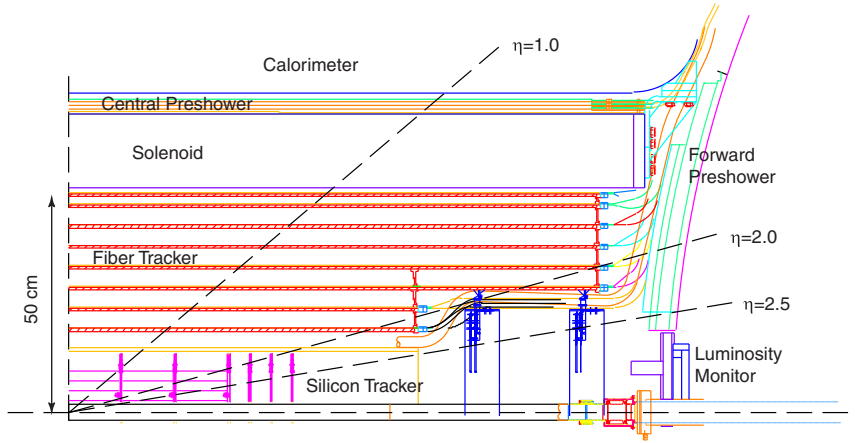} % postscript image file name
%\epsfbox{d0upgrade.eps} % postscript image file name
\caption{$r-z$ view of the \d0 tracking system. }

\end{figure}

\subsection{Central Fiber Tracker (CFT)}
The detector just outside the SMT is the 8-layered CFT, which is based on
scintillating fiber ribbon doublets with visible light photon counter (VLPC)
readout~\cite{vlpc}. Each layer contains 2 fiber doublets in a $zu$ or $zv$ configuration ($z$ = axial
fibers and $u,v = \pm 3^\circ$ stereo fibers). Each doublet
consists of two layers of 830~$\mu$m diameter fibers with 870~$\mu$m
spacing, offset by half the fiber spacing. This configuration provides very
good efficiency
and pattern recognition and results in a position resolution of
$\approx$~100~$\mu$m in $r \phi$. There are a total of 77,000 channels.

The CFT serves two main functions. First, with the SMT it enables track reconstruction
and momentum measurement for $\eta$~$<$~$\pm$1.7, second it provides fast Level 1 triggering on charged track momentum.

The fibers are up to 2.5~m long and the light is piped out by clear
fibers of length 7-11~m to the VLPCs situated in cryostats outside the
tracking volume, which are maintained at 9$^\circ$K.  The VLPCs are
solid state devices with a pixel size of 1~mm, matched to the fiber
diameter. The fast risetime, high gain and excellent quantum
efficiency of these devices make them ideally suited to this
application.

\subsection{Superconducting Solenoid}

The momenta of the charged particles will be determined from their curvature
in the 2T magnetic field provided by a 2.8 m long, 1.42 m in diameter and 1.1
 radiation length thick solenoid magnet. From the value of
$\sin\theta\times\int B_z dl$ and the space point precision provided
by the silicon and fiber tracking
system (located inside the solenoid) a momentum resolution of $\delta p_T / p_T = 0.002 p_T$, ($p_T$ in GeV) is
 expected.

\subsection{Preshower detectors}
The central and forward preshower detectors (CPS and FPS) are based on a
similar technology, using triangular scintillator strips (axial and 20$^\circ$
 \mbox{stereo}) with wavelength
shifter readout. These detectors provide fast energy and position measurements
for the electron trigger and facilitate offline electromagnetic identification.

\section{Muon Detectors}

The higher event rates in Run II have led us to add new muon trigger detectors
 covering full pseudorapidity range and the harsh radiation environment has
prompted us to replace the forward proportional drift tubes(PDTs) with
mini-drift tubes (MDTs). In the central region, the Run I PDTs are used
 and the front-end electronics have been replaced to ensure
dead-timeless operation.

The scintillation counters provide the time information and match
the muon tracks in the fiber tracker, and consists of three layers
to reduce hit combinatorics. The drift tubes provide enhanced muon
momentum resolution and pattern recognition. The design of the muon
system reduces backgrounds and trigger rates with additional shielding
and we have the ability to trigger on inclusive single muons with
$p_T > 7$ GeV and dimuons with $p_T > 2$ GeV.

\section{Trigger and Data Acquisition}
The \d0 trigger and DAQ systems have been completely restructured to
handle the shorter bunch spacing and new detector subsystems in
Run~II.  The Level 1 and 2 triggers utilize information from the
calorimeter, preshower detectors, central fiber tracker, and muon
detectors. The Level 1 trigger reduces the event rate from 7.5~MHz to
10~KHz and has a latency of 4~$\mu$sec. The trigger information is
refined at Level 2 using calorimeter clustering and detailed matching
of objects from different subdetectors. The Level 2 trigger has an
output rate of 1~KHz and a latency of 100~$\mu$sec. Level 3,
consisting of an array of PC processors, partially reconstructs event
data within 50~msec to reduce the rate to 50~Hz. Events are then
written to tape.

\begin{figure}[t]
%\figurebox{20pc}{15pc}{} % to have a box alone
\epsfxsize=27pc % will enlarge or reduce the postscript figures based on the xsize
\epsfbox{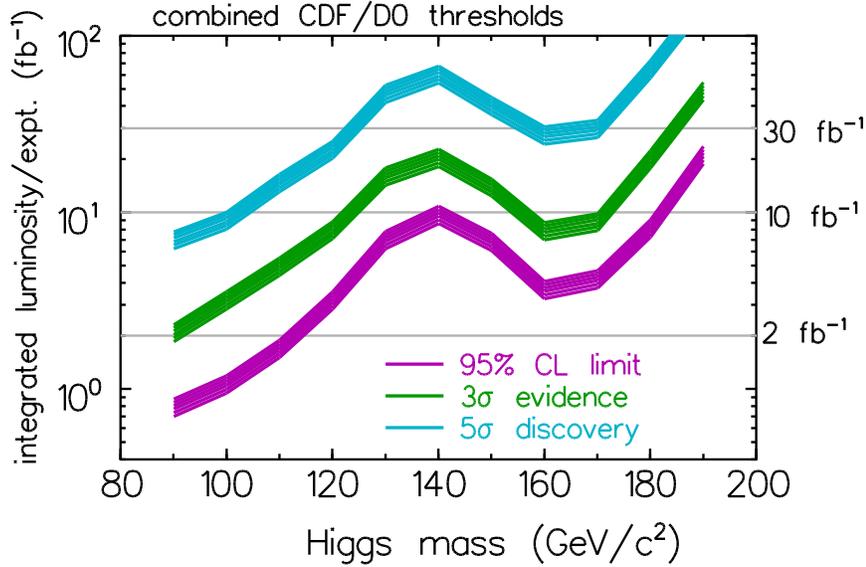} % postscript image file name
%\epsfbox{d0upgrade.eps} % postscript image file name
\caption{Integrated luminosity required as a function of Higgs mass for a 95\% C.L.
exclusion, 3$\sigma$ evidence and 5$\sigma$ discovery.}

\end{figure}

\section{Run II Physics Prospects - Some Highlights}
We have seen that the design of the Run II \d0 detector makes important
additions to the \d0 physics capabilities, namely: energy/momentum matching for electron identification, improved muon momentum resolution, charged sign and momentum determination, calorimeter calibration and displaced vertex identification (b-tags).

One of the crucial physics goals of Run II is to search for the Higgs boson.
In Run II the reach for Higgs at the Tevatron suggests a possibility of
discovering light Higgs, given sufficient integrated luminosity from both
experiments (\d0 and Collider Detector at Fermilab, CDF)
at the collider~\cite{run2_higgs}. The dominant decays can be classified into $M_ {Higgs} < 135~GeV$ and $M_{Higgs} > 135~GeV$. The light mass range is of great interest, since minimal
supersymmetric versions of SM always predict a light Higgs boson lying in this region. If there is no SM Higgs, 95\% confidence level exclusion limits can be set up to $\approx$120 GeV mass with 2~fb$^{-1}$. However if there is a Higgs, discovery at the 5$\sigma$ level can be obtained with 20~fb$^{-1}$ up to similar mass. This sensitivity is shown in \mbox{Figure 3}.

A detailed study of the top quark is another important physics interest.
 Since the CM energy is expected to increase from
1.8~TeV to $\approx$2.0~TeV, the $t \bar t$ production cross section will
increase by about 38\% (the production is dominated by $q \bar q \to t
\bar t$). For single top production the increase will be about 22\%
for the $s$-channel ($q \bar q \to t \bar b$), and about 44\% for the
$W$-gluon fusion process ($qg \to q t \bar b$). The expected
uncertainty in the top quark mass measurement is $\delta m_t \simeq
2$~GeV, and we should also be able to observe single top production for the first time.

SM has been very successful so far and in the electroweak sector, the goal is to make more precise tests of the SM and search for deviations that may signal the presence of new physics. A prime measurement is of the W mass, which
in Run II would be done with a precision of ~30 MeV from both \d0 and CDF~\cite{run2_mw}. Within SM, the mass of the top quark and the W boson set constraints on the mass of the Higgs boson.

In physics beyond the SM, there is potential for observing supersymmetry. For example, the supersymmetric  gaugino pair production in the trilepton
decay modes $p \bar p \to \tilde \chi _1^\pm \tilde \chi_1^\mp,
\tilde \chi _1^\pm \tilde \chi_2^0 \to 3\ell + X$, can be discovered with a mass reach of 220
GeV. Supersymmetric squark/gluino production in the $\mathrm{jets} + \etmis$
channel is expected to probe masses up to about 400~GeV for
2~fb$^{-1}$. Other interesting searches include a wide range of topics, namely,
leptoquarks, new gauge bosons ($W^\prime$, $Z^\prime$), topcolor, compositeness, technicolor and
extra dimensions. In Run II there is an excellent chance to discover new physics or exclude significant regions of parameter space.

A wide range of b-physics and QCD physics will also be pursued, but their details are beyond the scope of this article. Details of the \d0 detector and its upgrade can be found elsewhere~\cite{upgrade_details}.

\section*{Acknowledgments}
I would like to thank the Lake Louise Winter Institute for arranging a \mbox{stimulating} set of talks and for their warm hospitality. I would also like to thank Darien Wood for careful reading of this paper.


\begin{thebibliography}{99}

\bibitem{svx2e}T.~Zimmerman {\it et al}, \Journal{\IEEE}{42} {803} {1995}.

\bibitem{vlpc}M.D. Petroff and M.G. Stapelbroek, \Journal{\IEEE}{36}{158}{1989};M.D. Petroff and M. Atac, \Journal{\IEEE}{36}{163}{1989}.

\bibitem{run2_higgs}M. Carena {\it et al}, Report of the Tevatron Higgs Working Group, hep-ph/0010338.

\bibitem{run2_mw}R. Brock {\it et al}, Report of the  Working Group on Precision Measurements, hep-ex/0011009.

\bibitem{upgrade_details}Report on ``The \d0 upgrade: The Detector and its Physics, Fermilab Pub-96/357-E; J. Ellison, ``The \d0 Detector Upgrade and Physics Program, hep-ex/0101048.

\end{thebibliography}
\end{document}